# DIPOLE-GLASS CONCEPT AND HISTORY-DEPENDENT PHENOMENA IN RELAXORS

P.N. Timonin

Southern Federal University

344090, Rostov-on-Don, Russia

*The possibility to explain basic physical properties of relaxors within the concept of the dipole-glass transition is discussed. We argue that this concept provides the only consistent picture accounting of all known anomalous features of relaxors. The origin of their history-dependent properties can be naturally traced to the main paradigm of glass-state theory – the existence of numerous metastable states. Based on this paradigm phenomenological description of known history-dependent phenomena in relaxors agrees qualitatively with experiments.*

Strongly disordered ferroelectrics such as ferroelectric solid solutions and ceramics can exhibit rather unusual 'relaxor' properties among which the most notorious is the extremely slow relaxation of polarization which had given birth to the 'relaxor ferroelectrics' notion. Yet the most important feature of this class of ferroelectric materials is that disorder completely destroys the ferroelectric transition in them. Thus in zero electric field in all temperature range down to $T = 0$ no spontaneous polarization or ferroelectric domains appear in relaxors as well as there are no changes in their (average) crystalline structure. So quite naturally the relaxors have just broad maximum in the temperature dependence of dielectric susceptibility instead of sharp peak and the position of this maximum shifts to lower $T$ at lower frequencies, see Fig.1.

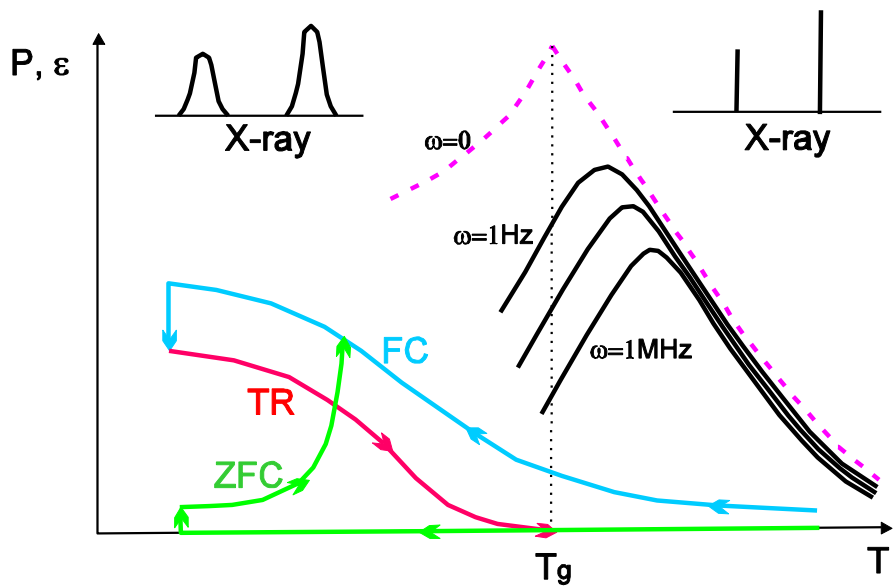

Fig. 1. Synopsis of basic physical properties of relaxors: frequency-dependent peak of $\varepsilon(T)$, appearance of thermoremanent (TR) polarization after field-cooling; different $P_{FC}$ and $P_{ZFC}$; broadening of Bragg peaks in X-ray diffraction.

In spite the apparent absence of ferroelectric transitions in relaxors the ferroelectric polarization can be induced in them at low $T$ via application of sufficiently strong external field for a limited period of time. Otherwise one can cool the sample in strong enough field to some low T and then switch off the field to find that it acquires some polarization which remains stable at the laboratory time scales, that is for hours and days at least. Subsequent heating of relaxor in zero field shows that such remanent polarization persists up to some critical $T_g$, see Fig.1. The appearance of stable remanent polarization in the material with no ferroelectric transition is a great paradox and inexplicable enigma for the classical theory of phase transitions. First there appeared rather naïve attempt to resolve this paradox using the notion of 'diffuse phase transition' in which different regions of a sample transform sequentially into ferroelectric phase throughout some temperature interval. This concept contradicts apparently to the X-ray diffraction experiments revealing no macroscopic polar regions in relaxors at all $T$ in zero field, see, for example, Ref. [1].

Much more consistent explanation of this enigma was advanced in the theoretical studies of the disordered magnets such as magnetic alloys which are the complete magnetic analogs of ferroelectric relaxors shearing with them all the features mentioned above. Namely it was suggested that such random systems has actually a sharp phase transition but in very specific phase nicknamed "spin-glass phase". In such phase the local spontaneous moments appear below some $T_g$ but their directions and magnitudes are chaotic which results in zero net moment of a sample. Such concept agrees perfectly with the experimental evidences of the absence of the (average) symmetry changes and macroscopic magnetized (polar) regions both in random magnets and relaxors. And it allows to explain why they become magnetized (polarized) under application of external field – it just align the local moments more or less parallel to its direction and they stay partially aligned after switching off the field. Naturally the degree of such field-induced order could depend on the strength of the applied field – the greater field the larger remanent order and remanent net moment.

The last supposition is actually the conclusion of the mean-field theory of spin-glasses [2]. It predicts the emergence below $T_g$ numerous long-lived (metastable) states caused by the so-called "frustration" induced by random interactions. This means the inability to satisfy the minimum (free)-energy condition by the finite set of local moments' configurations; instead the system with random interactions may have a huge number of such configurations with the same (or nearly the same) minimal energy or free energy (at $T\neq0$). This implies that the free energy of such random system considered as a function of some suitable physical quantities has a huge number of local minima. In other words, its free energy landscape is extremely rugged. Accordingly the relaxation of such system into some most deep local minimum may take a very long macroscopic time needed to overcome numerous free energy barriers.

Thus borrowing from the spin-glass theory the concept of the transition into the glassy state we can explain qualitatively the basic features of relaxors. So we may consider them as the systems experiencing the transition into "dipole-glass phase" with chaotic local spontaneous moments. And if this is do the case realized in relaxors one can easily find experimentally the transition point $T_g$ where such phase appears. We should only notice that all field - induced polar states should loose their remanent polarization on heating through this temperature, cf. Fig.1. Moreover, extrapolating the temperature dependencies of $\varepsilon(\omega)$ to $\omega = 0$ we would find that the limiting static $\varepsilon(0)$ has a break at the same $T_g$, see Fig.1.

Here we should stress that the existence of such unique $T_g$ is the very essence of the glass transition concept and should be the universal feature of all relaxor compounds irrespective of the nature of disorder inducing their relaxor properties. The assumption of such universality is mostly based on the simple fact that to date we have no other concept able to consistently describe numerous peculiarities of relaxors or random magnets. Yet this universality still needs the experimental verification in the variety of existing relaxors. Thus now we can only mention the evidence of simultaneous depolarization of all polar metastable states near the same $T_g$ in pyrochlore relaxor ferroelectric $Cd_2Nb_2O_7$ [3].

Another universal feature of relaxors according to the dipole-glass concept is the history-dependence of their state and, hence, their thermodynamic parameters. The simple experimental demonstration of the history-dependence in relaxors is the comparison of field-cooled (FC) and zero-field-cooled (ZFC) polarizations, see Fig.1, and their difference is often used as indicator of the onset of glassy state [2]. Indeed, under field and temperature variations the rugged free-energy landscape can experience drastic qualitative changes. Some minima may vanish and other ones may appear thus compelling the system to travel through the changing landscape in search of new stable minima. Generally the result of such search would depend on the way in the field-temperature (*E-T*) plane contrary to the case when our system has just one free-energy minimum. In the last case it always follows this unique minimum and the final state depends only on the final *E* and *T*.

This property of the systems with multiple free-energy minima is called "nonergodicity" and actually the existence of just two minima is sufficient to make the system nonergodic. The simple example is uniaxial single-domain ferroelectric having below its transition point two stable states with positive and negative polarization *P* in the region of *E-T* plane where module *E* is less than coercive field $E_c$, $|E| < E_c(T)$, see Fig.2a. If we cool it to some low $T_0$ in positive field $E_0 < E_c$ it always arrives to the state with $P > 0$. But if we first cool it to the same $T_0$ in some negative field and then change isothermally the field to the same positive $E_0$ value we would have the state with $P < 0$ in the same point $(T_0, E_0)$. Apparently this history-dependence arises because we arrive on these two routes to different branches of the ordinary hysteresis loop, see Fig.2b. This effect can be made quite clear if we consider $P(E,T)$ surface resulting from the temperature evolution of hysteresis loop, see Fig.2c. This surface has two overlapping sheets over the region $|E| < E_c(T)$ and the above routes bring the system to upper and lower sheet correspondingly.

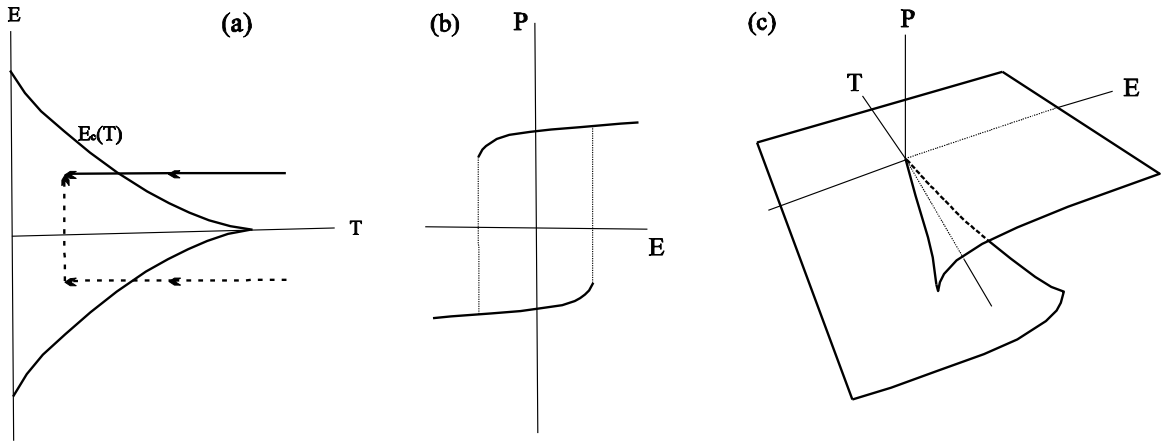

Fig.2. (a) The region of nonergodicity $|E| < E_c(T)$ of single-domain ferroelectric where different paths can lead to the different states on the upper or lower branches of hysteresis loop (b). (c) $P(E,T)$ surface formed by the evolving hysteresis loop.

The mechanism of history-dependence in relaxors is essentially the same but a bit more complicated owing to huge number of stationary (metastable) states with different degree of order and the whole range of polarizations. Apparently, each metastable state would have its separate $P(E,T)$ sheet and we would have the whole stack of such sheets in relaxors. How this stack can look like? To answer this question we may recall that relaxors also have hysteresis loops [4]. Mostly their form differs from that of ordinary ferroelectric one in Fig.2b – usually they are inclined. After a little thought we may conclude that the only place for the stack of metastable $P(E,T)$ sheets is the interior of such expanding at lower T inclined loop, see Fig.3a. Apparently, at some fixed $T_0$ we would have the inclined loop densely filled with metastable $P(E,T_0)$ curves which end up at its sides, see Fig.3b. The cross-section of our stack with plane $E = E_0$ gives the temperature dependencies of polarizations $P(E_0,T)$ in different metastable states, which look like those in Fig.3c.

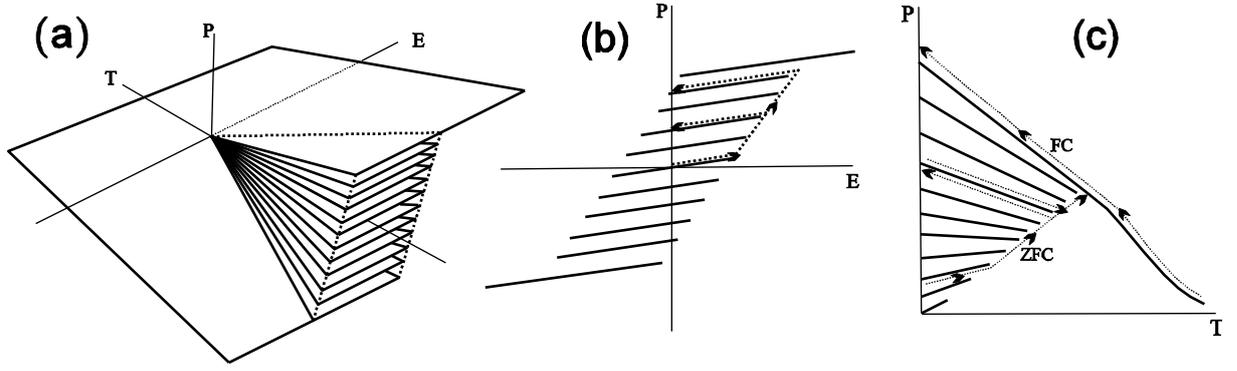

Fig.3. Schematic picture of metastable *P(E,T)* sheets (a) and its cross-sections $T=T_0$ (b) and $E=E_0$ (c). Directed lines show possible variations of P in different processes.

These qualitative considerations on the form of metastable *P(E,T)* sheets and their resulting cross-sections (Figs. 3b, 3c) provide us with the clear notion of the history-dependent outcome in quasi-static processes in relaxors. Term "quasi-static" means here that temperature and field variations are sufficiently slow for system to be able to follow the changing minimum until it vanishes. Further we assume that leaving the vanishing minimum a relaxor goes to that having nearest polarization value. With these assumptions we can easily see from Fig. 3b why the temporary application of field brings relaxor into a polarized state and why there is the threshold field for this process. As well the temperature dependencies of metastable polarizations in Fig. 3c elucidate the origin of different polarizations in FC and ZFC processes. Moreover Fig. 3c predicts the appearance of multiple polarizations curves under cyclic heating-cooling process.

To gain full command over quasi-static history-dependent processes in relaxors we need also the way to describe the thermodynamic properties of the individual metastable states. This can be achieved in the framework of Landau-type phenomenology using the notion of the multiple polar modes' condensation [5, 6]. The Landau potential of a generic uniaxial relaxor depending on the macroscopic number of polarizations $P_i$, $i=1,…,N_0$, has a form

$$F = \frac{\tau_g}{2}\left([P^2]-[P]^2\right) + \frac{\tau_f}{2}[P]^2 + \frac{1}{4}[P^4] + \frac{b}{4}[P^2]^2 + \frac{c}{2}[P^2][P]^2 + \frac{d}{4}[P]^4 - E[P],$$

$$[P^k] \equiv \frac{1}{N_0}\sum_{i=1}^{N_0} P_i^k$$

It describes the competition between ferroelectric order, $[P^2]=[P]^2 \neq 0$, and dipole-glass one, $[P]=0$, $[P^2] \neq 0$, which can appear at $\tau_f < 0$ and $\tau_g < 0$ correspondingly. The potential *F* is symmetric under $P_i$ permutations thus imitating the frustration-induced degeneracy of stable states. It can have up to $2^{N_0}$ local metastable minima in the dipole glass phase and, quite remarkably, their $P(E,T)$ sheets form the interior of inclined hysteresis loop in this phase in accordance with the above qualitative considerations. Each metastable state can be characterized by the fraction $n$ of positive $P_i$ (up to their permutations).

Analysis of possible equilibrium phases is reduced to the finding the global minimum of $F$. It is comparatively simple when $b, c, d \geq 0$. Then we have for $d(1+b) < c^2$ just two phases (apart from paraelectric): dipole-glass (DG) at $\tau_g < 0$, $\tau_g < \tau_f$ and ferroelectric one ($F$) at $\tau_f < 0$, $\tau_f < \tau_g$ with the equilibrium first order transition between them at $\tau_g = \tau_f$. When $d(1+b) > c^2$ there are three phases: DG at $\tau_g < 0$, $\tau_g\left(1+\dfrac{c}{1+b}\right) < \tau_f$, partially polarized mixed phase (M) with $[P^2] \neq [P]^2 \neq 0$ at $\tau_g\left(1+\dfrac{c+d}{1+b+c}\right) < \tau_f < \tau_g\left(1+\dfrac{c}{1+b}\right) < 0$ and F-phase at $\tau_f < 0$, $\tau_f < \tau_g\left(1+\dfrac{c+d}{1+b+c}\right)$. The transitions between them are also of first order. Yet the equilibrium properties of relaxors are generally non-observable due to their rugged landscape preventing the system to relax into the global minimum on the laboratory times. They are relevant as much as they show up in the unavoidably nonequilibrium real-time experiments. Thus some signatures of SG-M-F transitions can be found in the changes of hysteresis loop form. As Fig. 4 show region of the M-phase roughly corresponds to that where vertical jumps develop at loop sides, while the F-phase mostly has vertical loop sides. Each curve inside the loop has its own parameter $n$ mentioned above so the upturns of $P_i$ with the changes of $n$ can take place at the sides of a loop only when system comes from one curve to another. This will result in the emergence of electromagnetic and acoustic noise when system sweeps the loop's sides. Note also that the possibility to observe numerous small loops inside the main one is the direct manifestation of the metastable states' multiplicity. The temperature evolution of the hysteresis loop similar to that in Fig. 4 was found in relaxor PMN-PT [7].

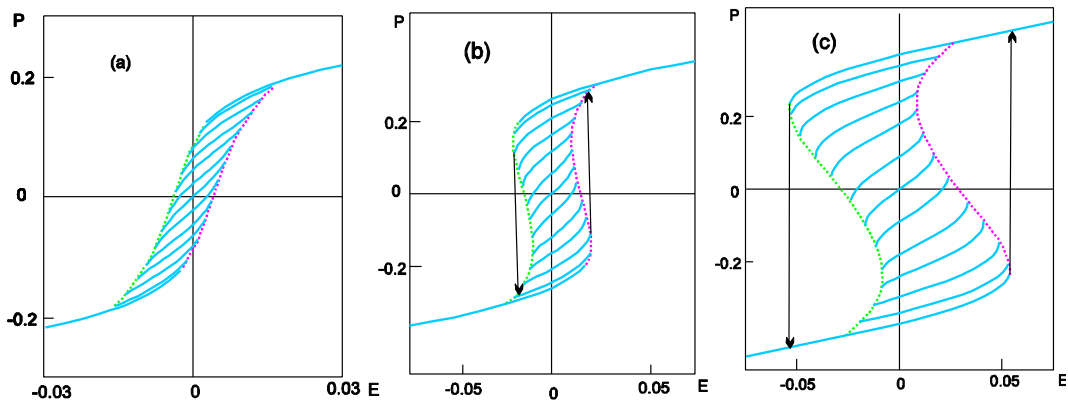

Fig.4. Evolution of the hysteresis loop in the model with $b=0.1$, $c=0.5$, $d=1$, $\tau_f = 1.5\tau_g + 0.1$. (a) DG phase, $\tau_g = -0.05$, (b) M phase, $\tau_g = -0.125$, (c) F phase, $\tau_g = -0.2$,

Meanwhile the anomalies in the temperature dependencies of polarization, dielectric permittivity or heat capacity have usually no relation to the equilibrium transitions. Generally they appear when relaxor comes to the stability limit of its present metastable state. The position and the very presence of such anomalies depend on the sample's previous history as Figs. 5, 6 show. Here $P(E_0,T)$ are obtained for different metastable states as cross-sections by the $E = E_0$ plane of the stack metastable $P(E,T)$ sheets. So the limits at which the individual curves terminate are the cross-sections of the loop's sides. Thus the smooth variation of $P(E_0,T)$ and $\varepsilon(E_0,T)$ continues toward the stability limit of a given state and then anomaly appears reflecting the transition to another curve. This can explain the additional anomaly of $\varepsilon(E_0,T)$ in FC process as compared to $\varepsilon(0,T)$ in ZFC one and the field-induced anomaly of $\varepsilon(E_0,T)$ in strong enough field in field-heating (FH after ZFC) process, cf. Figs. 5, 6. Such anomalies albeit with smaller amplitude were observed in PMN [8]. Some discrepancies with present theory are unavoidable if the quasi static regime is violated or the high-frequency $\varepsilon(E_0,T)$ is measured.

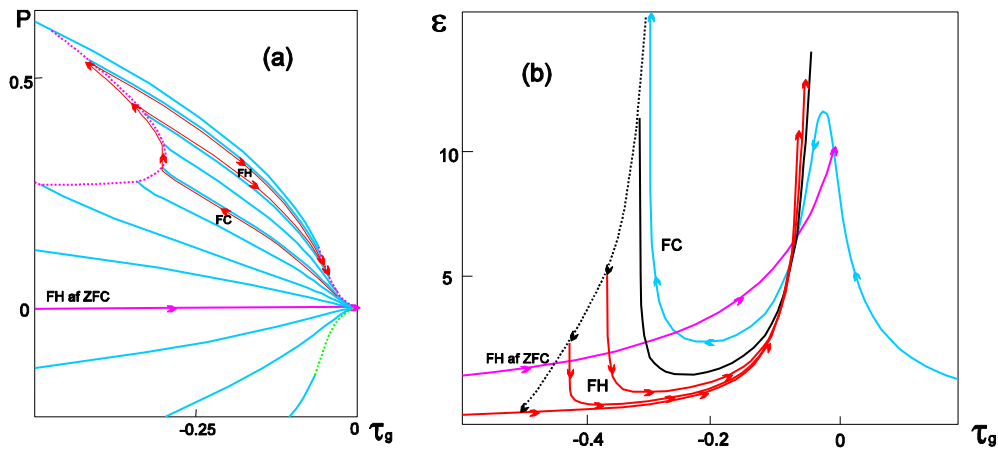

Fig.5. Temperature dependencies of $P$ (a) and $\varepsilon$ (b) of relaxor in various metastable states in small field $E = 3 \times 10^{-4}$. Parameters of $F$ are the same as in Fig.4. Directed lines show the evolution of $P$ and $\varepsilon$ in different regimes of $E$ and $T$ variation.

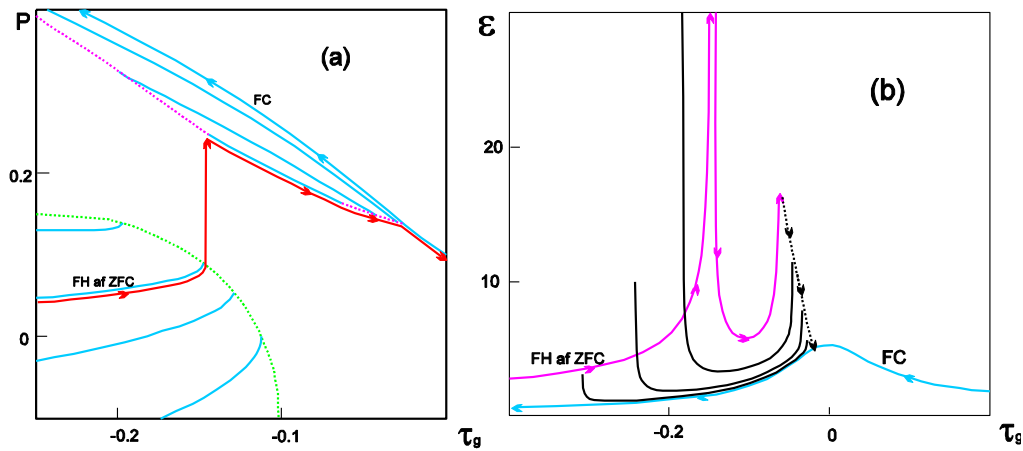

Fig.6. The same as in Fig. 5 for the strong field $E = 0.013$.

Thus we see that the history-dependent phenomena in relaxors can be quite naturally explained by the existence of the numerous metastable sheets in the interior of the hysteresis loop. Temperature variations of loop form can compel the system to leave one sheet and to join another thus causing relaxor to exhibit history-dependent anomalies. So the adequate understanding of the physical properties of relaxors implies the thorough studies of the forms of the metastable sheets, its boundaries and the thermodynamic properties of relaxors on them. It is important to realize that all metastable states inside the loop enjoy equal rights the same as different thermodynamic phases in which the system can stay very long (probably infinitely long). To date we are very far from the comprehensive investigations of this sort. Yet this way seems to be very promising in view of the universality of the dipole-glass concept and the possibility to gain the full control on the elusive nature of relaxors. Apart from the realization of the simple qualitative ideas behind the glass concept this needs also some revisions in the experimental technique. Namely, to achieve the quasi-static regime sufficiently low rates of $T$ and $E$ variations should be used. It seems that for the PMN – type relaxors cooling (heating) rates should not exceed 0.1 K/min and ac-field frequencies should be less then 0.01 Hz. Quite probably with these prerequisites the reproducible quasi-static results can be achieved making the concept of glassy states in relaxors firmly justified.

REFERENCES

1. C. Stock, Guangyong Xu, P. M. Gehring, H. Luo, X. Zhao, H. Cao, J. F. Li, D. Viehland, and G. Shirane, Neutron and x-ray diffraction study of cubic [111] field-cooled Pb(Mg$_{1/3}$Nb$_{2/3}$)O$_3$. Phys. Rev. B, **76**, 064122 (2007).

2. K. Binder and A. P. Young, Spin glasses, Rev. Mod. Phys. **58**, 801(1986).

3. N. N. Kolpakova, P. Czarnecki, W. Nawrocik, M. P. Shcheglov, P. P. Syrnikov and L. Szczepanska, Crossover from glassy to ferroelectric polarization behavior under a dc bias electric field in relaxor ferroelectrics, Phys. Rev. B, **72**, 024101 (2005)

4. X. Zhao, W. Qu, X. Tan, A. A. Bokov and Z.-G. Ye, Electric field-induced phase transitions in (111)-, (110)-, and (100)-oriented Pb(Mg$_{1/3}$Nb$_{2/3}$)O$_3$ single crystals. Phys. Rev. B, **75**, 104106 (2007).

5. P. N. Timonin, Nonergodic thermodynamics of disordered ferromagnets and ferroelectrics. Eur. Phys. J. B, **64**, 125 (2008).

6. P. N. Timonin, Phenomenological theory of nonergodic phenomena in dipole- and spin-glasses. Eur. Phys. J. B, **70**, 201 (2009).

7. S. M. Emelyanov, F. I. Savenko, Yu. A. Trusov, V. I. Torgashev and P. N. Timonin, Dilute ferroelectric in random electric field: phase transitions in PMN-PT crystals, Phase Transitions, **45**, 251 (1993).

8. Zuo-Guang Ye and Hans Schmid, Optical, dielectric and polarization studies of the electric-field-induced phase transitions in PMN, Ferroelectrics, **145**, 83 (1993).